# The Hidden Cost of Using Amazon Mechanical Turk for Research

Antonios Saravanos[1(✉)[0000-0002-6745-810X]], Stavros Zervoudakis[1], Dongnanzi Zheng[1], Neil Stott[2], Bohdan Hawryluk[1], and Donatella Delfino[1]

[1] New York University, New York, NY, 10003, USA
[2] Cambridge Judge Business School, Cambridge, Cambridgeshire, CB2 1AG, UK
{saravanos, zervoudakis, dz40}@nyu.edu, n.stott@jbs.cam.ac.uk, {bh54, dd61}@nyu.edu

**Abstract.** In this study, we investigate the attentiveness exhibited by participants sourced through Amazon Mechanical Turk (MTurk), thereby discovering a significant level of inattentiveness amongst the platform's top crowd workers (those classified as 'Master', with an 'Approval Rate' of 98% or more, and a 'Number of HITS approved' value of 1,000 or more). A total of 564 individuals from the United States participated in our experiment. They were asked to read a vignette outlining one of four hypothetical technology products and then complete a related survey. Three forms of attention check (logic, honesty, and time) were used to assess attentiveness. Through this experiment we determined that a total of 126 (22.3%) participants failed at least one of the three forms of attention check, with most (94) failing the honesty check – followed by the logic check (31), and the time check (27). Thus, we established that significant levels of inattentiveness exist even among the most elite MTurk workers. The study concludes by reaffirming the need for multiple forms of carefully crafted attention checks, irrespective of whether participant quality is presumed to be high according to MTurk criteria such as 'Master', 'Approval Rate', and 'Number of HITS approved'. Furthermore, we propose that researchers adjust their proposals to account for the effort and costs required to address participant inattentiveness.

## 1 Introduction

Over time, online services for participant recruitment by researchers have increased in popularity [30]. Amazon Mechanical Turk (MTurk; also known as Mechanical Turk [15]) is one of the oldest and most frequently selected tools from a spectrum of web-based resources, enabling researchers to recruit participants online and lowering the required time, effort, and cost [24, 39]. A Google Scholar search for the term 'Mechanical Turk' reveals continuing growth in its use, with 1,080, 2,750, and 5,520 items found when filtering the results for 2010, 2012, and 2014 respectively [24]. The technology facilitates "an online labor market" where "individuals and organizations (requestors)" can "hire humans (workers) to complete various computer-based tasks", which they describe as "Human Intelligence Tasks or HITs" [24]. The MTurk platform's suitability for use in research has been extensively evaluated [12], with most published studies describing it as a suitable means for recruitment [8], although a few also state reservations [9]. This paper builds on that work by investigating the reliability of the top crowd workers that can potentially be sourced from the MTurk platform, while concurrently motivating them through offers of high compensation. Specifically, we focus on the attentiveness exhibited by US based workers classified as 'Master', who had completed at least 98% of the tasks they committed to completing (i.e., an 'Approval Rate' of 98% or more). Additionally, the number of these activities was required to exceed 999 (i.e., the 'Number of HITS approved' had a value of 1,000 or more). To the best of our knowledge, this segment of the platform has yet to be studied with a focus on participant attention.

# 2 Background

Amazon does not disclose real-time data on the total number of workers available for hire via their MTurk service, or those online at any particular moment. Several researchers offer insight into what those values might be [9, 13, 36]. Ross et al. [36] report that in 2010 the platform had more than 400,000 workers registered. Likewise, there were anywhere between 50,000 to 100,000 HITs at any given time. In 2015, Chandler, Mueller, and Paolacci [9] wrote: "From a requester's perspective, the pool of available workers can seem limitless, and Amazon declares that the MTurk workforce exceeds 500,000 users". Stewart et al. [40] report that the turnover rate is not dissimilar to what one would experience in a university environment, with approximately 26% of the potential participants on MTurk retiring and being replenished by new people. More recently (2018), Difallah, Filatova, and Ipeirotis [13] found that at least 100,000 workers were registered on the platform, with 2,000 active at any given time. The authors also state that a significant worker turnover exists, with the half-life for workers estimated to be between 12 and 18 months [13]. Such numbers as those reported by Stewart et al. [40] and Difallah, Filatova, and Ipeirotis [13] indicate that recruiting the same worker more than once for a given experiment is highly unlikely.

MTurk is probably the most thoroughly studied of the available platforms for online participant recruitment through crowdsourcing. The literature on the suitability of MTurk for research presents a somewhat 'rosy' picture, labeling it as adequate for use with experiments. This includes the work of Casler, Bickel, and Hackett [8], who compared the data obtained through the recruitment of participants on MTurk with data collected from participants recruited through social media, and those recruited on an academic campus [8]. The authors found that the data was similar across all three pools and highlight that the MTurk sample was the most diverse [8]. Moreover, the authors [8] reveal that the results were similar irrespective of whether the experiments had been completed in-lab or online. Both the replicability and reliability of data collected through the MTurk platform have been established. Rand [34] found that participant responses across experiments were consistent, allowing for replication of results and Paolacci, Chandler, and Ipeirotis [31] found increased reliability of data. Hauser and Schwarz [17] found that participants recruited from MTurk exhibited superior attention to the assigned task compared to participants recruited using traditional approaches.

Despite all these reassuring findings, a growing body of literature raises warnings that must be addressed [2, 5, 20, 25, 50], and "significant concerns" remain [15]. Such concerns are not new. For example, the use of attention checks to identify inattentive participants was standard practice for a large group of research communities, and prior work from past decades shows that many participants (from 5% to 60%) answer survey questions carelessly [6, 21, 28]. However, to some extent, this would not be expected with MTurk, as it would be assumed that individuals are essentially workers, and as such, they would be devoted to the task and paying attention. The main concern is the assertion that the low remuneration attracts workers with limited abilities who cannot find better employment [15]. Stone et al. [42] note that participants recruited through MTurk "tend to be less satisfied with their lives than more nationally representative samples", although they comment that "the reasons for this discrepancy and its implications are far from obvious". The reliability of crowd workers has been widely discussed and studied by investigating the impact of attentiveness on the reliability of the crowd worker responses (e.g., [37]).

Researchers are increasingly concerned that participants sourced through MTurk "do not pay sufficient attention to study materials" [15]. A prominent example is the work of Chandler et al. [9], who reveal that participants were not always entirely devoted to the assigned task and were instead multitasking. Litman et al. [25] identified that the practice of multitasking while participating in a research study is problematic, as it can lead to inattentiveness and reduce participants' focus on details [25]. Consequently, studies relying on participants devoting their full attention to the current work are at risk [25]. The authors also state that "these findings are especially troubling, considering that the participants in the Chandler et al. study were some of the most reliable of MTurk respondents, with cumulative approval ratings over 95%". The current research seek to understand this inattentiveness, trusting that our research will be useful to others who source participants for research using this tool.

## 3 Methodology

This section outlines the methodology used for the study.

### 3.1 Experimental Design

To investigate participant attention an experimental approach was adopted. Participants solicited through MTurk were forwarded to the Qualtrics web-based software where they were randomly presented with a vignette describing one of four hypothetical technology products. Subsequently, they were asked questions on their intention to adopt that technology. MTurk has been used on numerous occasions to understand user intention to adopt technology [29, 38, 41, 48, 49, 51]. Participants were then given one of two technology acceptance questionnaires to share their perceptions of the technology presented in their respective vignettes. The questionnaires were adap-tations of the most popular models used to study user adoption of technology [35]. The first questionnaire (short) reflected the instrument for the second version of the unified theory of acceptance and use of technology (UTAUT2) [45] model and comprised 52 questions in total. The second questionnaire (long) reflected the instrument for the third version of the technology acceptance model (TAM3) [46] and comprised 74 questions. Both questionnaires also included 10 demographic and experience questions. Aside from the demographic questions, each question was rated through a 7-point Likert scale ranging from 'strongly disagree' to 'strongly agree'.

**Assessing Attention.**
Three forms of attention check, derived from the work of Abbey and Meloy [1], were used to gauge participant attention. The first was a logical check based on logical statements. It required participants to demonstrate "comprehension of logical relationships" [1]. An example of such a question might be 'at some point in my life, I have had to consume water in some form'. This check comprised two such logical statements to answer, as shown in Table 1. The second was an honesty check to "ask a respondent directly to reveal their perceptions of their effort and data validity for the study" [1]. An example was, 'I expended effort and attention sufficient to warrant using my responses for this research study'. As part of this check, participants were asked two questions regarding their perception of the attention invested in the experiment. Table 1 shows the questions used. These questions were also rated using a 7-point Likert scale ranging from 'strongly disagree' to 'strongly agree'. Participants who did not respond to both questions by selecting the 'strongly agree' choice were deemed to have failed their respective attention checks.

**Table 1.** Attention Check Questions.

| Code | Type | Original Question | Adapted Question |
|---|---|---|---|
| ATT1 | Logical Statement | I would rather eat a piece of fruit than a piece of paper. | I would rather eat a piece of fruit than a piece of paper. |
| ATT2 | | At some point in my life, I have had to consume water in some form. | At some point in my life, I have had to consume water in some form. |
| ATT3 | Honesty Check | On a scale of 1-10, with one being the least attention and 10 being the most attention, please indicate how much attention you applied while completing this study. | I applied sufficient attention while completing this study. |
| ATT4 | | Did you expend effort and attention sufficient to warrant using your responses for this research study? | I expended effort and attention sufficient to warrant using my responses for this research study. |

The third form of attention check was a time check, which used "response time" to ascertain attention, employing the concept that response times might be "overly fast or slow based on distributional or expected timing outcomes" [1]. Participants who were unable to complete the experiment within a reasonable time were deemed to have failed that check. To estimate the response time, we totaled up the number of words that participants would read as part of the informed consent document, instructions, and vignette. We then used the conservative reading rate (200 words per minute) described by Holland [18] to estimate the time participants would require to read that material. To determine the time participants would need to complete each Likert question, we used the estimate provided by Versta Research [47], which is 7.5 seconds on average. Table 2 summarizes our calculations. Participants who spent less than 70% of the estimated time on the survey were deemed to have failed the time check.

**Table 2.** Task Composition and Participant Estimated Effort and Compensation.

| Group | Vignette | Consent (# of words) | Vignette (# of words) | # of Likert Questions | Participant Total Compensation | Estimated Completion Time (in seconds) | Effective Wage (in USD) |
|---|---|---|---|---|---|---|---|
| 1 | 1 | 265 | 80 | 52 | $3.41 | 494 | $24.85/hr |
| 2 | 2 | 265 | 97 | 52 | $3.41 | 499 | $24.60/hr |
| 3 | 3 | 265 | 106 | 52 | $3.41 | 501 | $24.50/hr |
| 4 | 4 | 265 | 124 | 52 | $3.41 | 507 | $24.21/hr |
| 5 | 1 | 265 | 80 | 74 | $4.19 | 659 | $22.89/hr |
| 6 | 2 | 265 | 97 | 74 | $4.19 | 664 | $22.72/hr |
| 7 | 3 | 265 | 106 | 74 | $4.19 | 666 | $22.65/hr |
| 8 | 4 | 265 | 124 | 74 | $4.19 | 672 | $22.45/hr |

**Compensation.**

A factor that was considered important and that needed to be controlled for was compensation. The concern was that the level of compensation might influence participant attention. However, numerous studies have investigated how compensation influences the quality of data produced by MTurk workers [4, 7, 25]. Most found that the quality of results is not linked to the rate of compensation, with Litman, Robinson, and Rosenzweig [25] stating that "payment rates have virtually no detectable influence on data quality". One example of such a study was conducted by Buhrmester et al. [7], who offered participants 2 cents (i.e., $0.25/hour), 10 cents, or 50 cents (i.e., $6 per hour) to complete a five-minute task. The authors found that while recruiting participants took longer when lower compensation was offered, the data quality was similar irrespective of the offered compensation. A similar study was conducted by Andersen and Lau [4], who provided participants with either $2, $4, $6, or $8 to complete a task. They found that the remuneration did not influence participants' performance, writing that there was "no consistent or clear evidence that pay rates influenced our subject behavior".

A smaller number of studies show that the quality of work produced by those on MTurk is influenced by the compensation size. An example is seen in Aker et al. [3] who compensated participants for a task at rates of $4, $8, and $10 per hour. Their "results indicate that in general higher payment is better when the aim is to obtain high quality results" [3]. Overall, most tasks on MTurk offer a minimal level of compensation [7, 14, 33]. In 2010, the mean and median wages were $3.63/hour and $1.38/hour, respectively [19]. In 2019, the median compensation in the United States was $3.01/hour [16]. Paolacci, Chandler, and Ipeirotis [31] comment that "given that Mechanical Turk workers are paid so little, one may wonder if they take experiments seriously".

The rate offered to participants in this study surpassed $22/hour in order to ensure that participants were adequately motivated and thereby control for compensation (Table 2). The size of this compensation could be considered excessive when considering what is traditionally offered to participants on MTurk, the federal minimum wage in the United States of $7.25/hour, and what is presented in studies examining the effect of compensation on performance. For example, Aker et al. [3] describe $10/hour as high. Offering an extremely generous wage was expected to negate undesirable effects and induce participants to devote their full attention to our study.

### 3.2 Participants

Participants were selected to represent the top workers the MTurk platform offers. This was accomplished by using a filtering mechanism, allowing only workers satisfying certain criteria to participate in the study [10]. The filters used were: 1) located in the United States; 2) classified by Amazon as 'Master' level; 3) had completed at least 98% of the tasks they committed to completing (i.e., an 'Approval Rate' of 98% or more); and 4) had completed at least 1,000 tasks (i.e., their 'Number of HITS approved' rating was 1,000 or more). The data was collected through two batches over 16 days (between December 22nd, 2019, and December 30th, 2019, and again between February 1st, 2020, and February 7th, 2020). Participation in this study was voluntary and all our participants were first asked to confirm that they were willing to participate before being allowed to begin the experiment. The privacy of participants was protected using confidential coding.

The sample was comprised of 564 participants, 293 (51.95%) identified as male, and 271 (48.05%) identified as female. Most participants were in the 31-55 age range (73.94%), had some form of undergraduate (74.47%) or postgraduate (9.93%) training, and earned below $60,000 (60.29%) per year. Most of the participants (480, equating to 85.11% of the sample) identified as white. Finally, most participants were either never married (284, or 50.35%) or married (215, or 38.12%). Table 3 shows the participants' demographics for the sample in greater detail. Figure 1 shows the locations of participants within the United States. All states were represented except for Wyoming, with the five most prevalent in the sample being California, Florida, Pennsylvania, Texas, and Michigan.

**Table 3.** Participant Demographics.

| Characteristic | Category | N | Percentage |
|---|---|---|---|
| Age | 18-25 | 18 | 3.19% |
| | 26-30 | 78 | 13.83% |
| | 31-55 | 417 | 73.94% |
| | 56 or older | 51 | 9.04% |
| Gender | Male | 293 | 51.95% |
| | Female | 271 | 48.05% |
| Income | Less than $10,000 | 17 | 3.01% |
| | $10,000 - $19,999 | 46 | 8.16% |
| | $20,000 - $29,999 | 70 | 12.41% |
| | $30,000 - $39,999 | 83 | 14.72% |
| | $40,000 - $49,999 | 63 | 11.17% |
| | $50,000 - $59,999 | 61 | 10.82% |
| | $60,000 - $69,999 | 59 | 10.46% |
| | $70,000 - $79,999 | 44 | 7.80% |
| | $80,000 - $89,999 | 24 | 4.26% |
| | $90,000 - $99,999 | 32 | 5.67% |
| | $100,000 - $149,999 | 47 | 8.33% |
| | $150,000 or more | 18 | 3.19% |
| Marital Status | Never Married | 284 | 50.36% |
| | Married | 215 | 38.12% |
| | Separated | 3 | 0.53% |
| | Divorced | 52 | 9.22% |
| | Widowed | 7 | 1.24% |
| | No response | 3 | 0.53% |
| Race | Asian | 32 | 5.67% |
| | Black or African American | 27 | 4.79% |
| | Other | 25 | 4.43% |
| | White | 480 | 85.11% |
| Schooling | < High school degree | 2 | 0.35% |
| | High school graduate | 85 | 15.07% |
| | Some college - no degree | 119 | 21.10% |
| | Associate’s degree | 72 | 12.77% |
| | Bachelor’s degree | 229 | 40.60% |
| | Master’s degree | 44 | 7.80% |
| | Professional degree | 8 | 1.42% |
| | Doctoral degree | 4 | 0.71% |
| | No response | 1 | 0.18% |

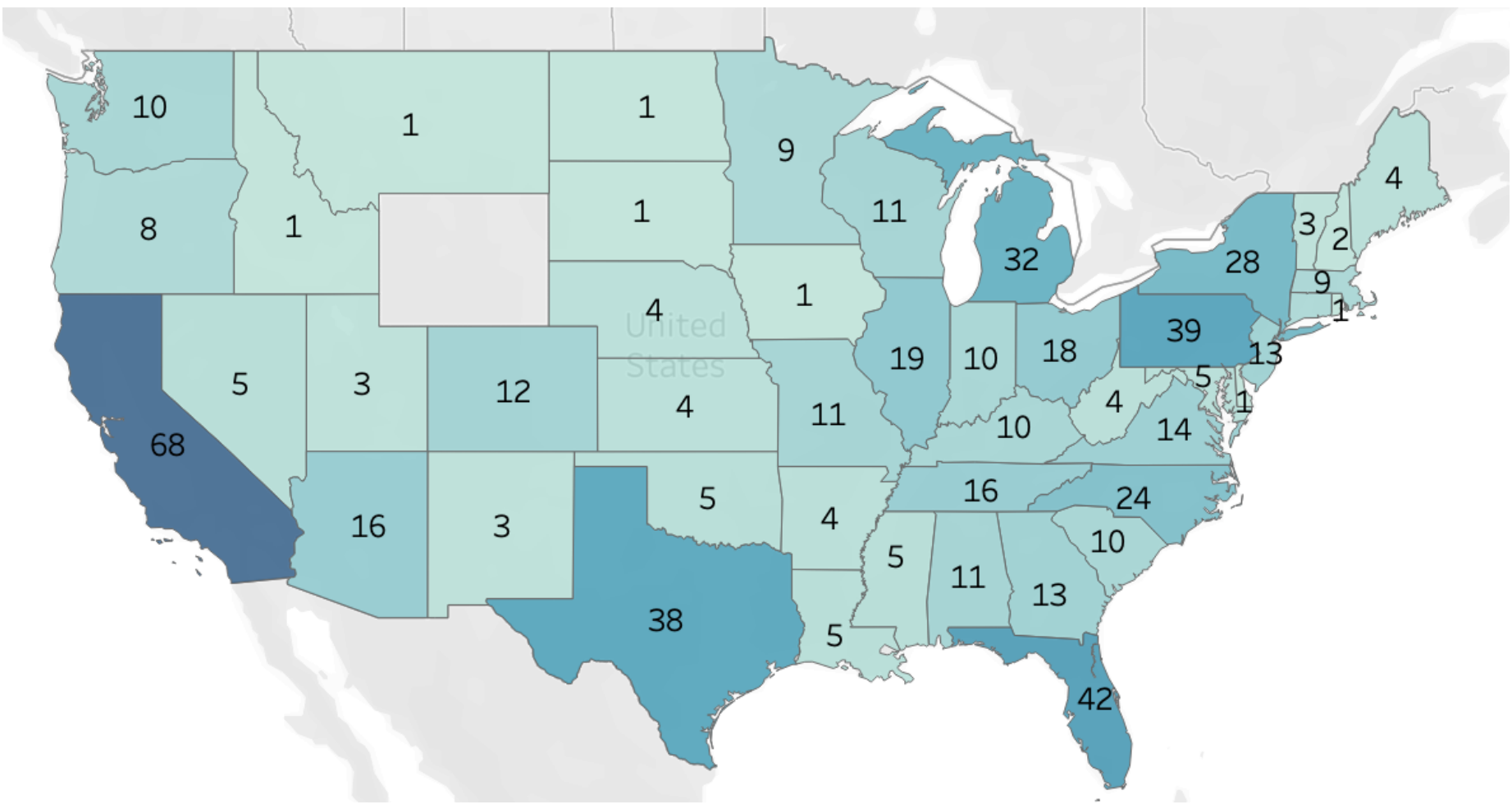

**Fig. 1.** Participants by State.

## 4 Analysis and Results

To analyze our data, we relied on three techniques. The first examined the frequency with which attention checks were passed or failed by participants; this revealed that 126 of the 564 participants (22.34%) failed at least one form of attention check. The attention check that most participants failed was the honesty check (94/564), followed by the logic check (31/564), and the time check (27/564). Some participants failed more than one check, with 14/564 (2.48%) failing both logic and honesty checks and 6/564 (1.06%) failing both time and honesty checks. Finally, 6/564 (1.06%) participants failed all three attention checks (logic, honesty, and time). Figure 2 illustrates the numbers of participants who failed and passed each form of attention check. As expected, participants who passed the time check were more likely to pass the other attention checks (logic and honesty).

The second technique used Spearman rank-order (rho) correlations to assess the correlation between the characteristics of age, gender, income, race, and prior experience with the technology and each of the three forms of attention checks (i.e., logic, honesty, and time checks). No significant correlation was found except in two instances. First, prior experience of using the technology being studied was positively correlated with the logic check ($r_s = 0.192$, $p = 0.000$). Second, prior experience of using the technology being studied was positively correlated with the honesty check ($r_s = 0.213$, $p = 0.000$). Therefore, the more familiar participants were with the technology, the more likely they were to pass the logic and honesty checks.

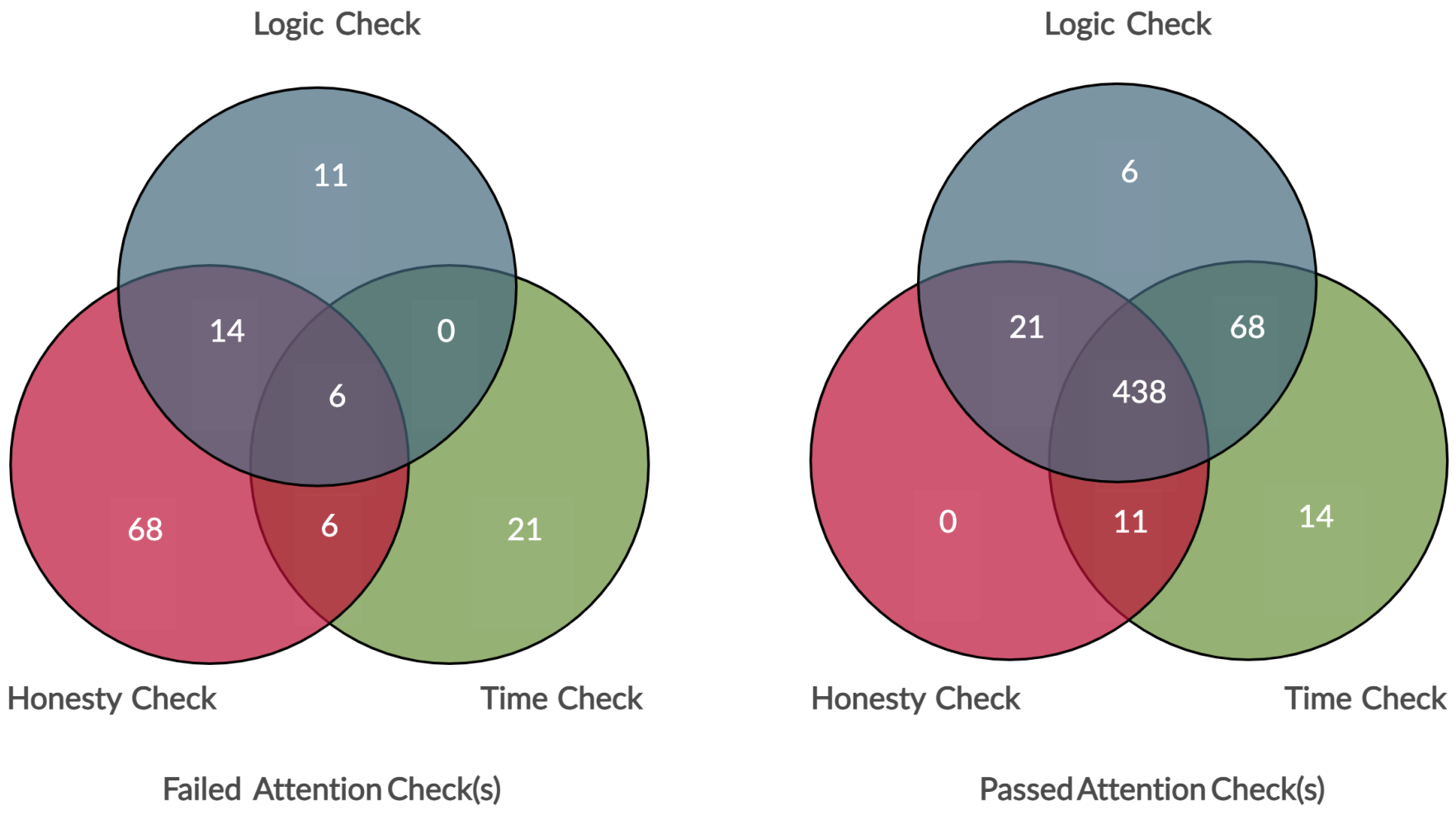

**Fig. 2.** Venn Diagram Depicting Participant Failure and Passing of Attention Check by Type.

Spearman rho correlations were also used to assess the relationship between the three different forms of attention checks and the passing of all three checks. A positive correlation was found between participants passing the logic check and passing the honesty check ($r_s = 0.310$, $p = 0.000$), failing the time check and failing the honesty check ($r_s = 0.132$, $p = 0.002$), and failing the time check and failing the logic check ($r_s = 0.139$, $p = 0.001$). That is, participants who pass one of the three attention checks are more likely to pass the other two attention checks. Table 4 shows the results of the correlations.

Finally, a logistic regression analysis was used to investigate whether age, income, gender, prior experience, and time on task influenced participant attention. All four assumptions required for logistic regression were satisfied [23]. Only prior experience with the technology in the logistic regression analysis contributed to the model (B = 0.484, SE = 0.136, Waid = 12.706, p = 0.000). The estimated odds ratio favored an increase of 62.3% [Exp(B) = 1.623, 95% CI (1.244, 2.118)] for participant attention for every unit increase in experience. None of the other variables were found to be statistically significant (Table 5).

**Table 4.** Nonparametric Correlations.

| Factor | AG | AT | EX | GE | GR | HC | IN | LC | MS | RA | SC | TC |
|---|---|---|---|---|---|---|---|---|---|---|---|---|
| Age (AG) | 1 | 0.034 | - 0.029 | - 0.133$^{**}$ | 0.030 | 0.036 | -0.017 | 0.035 | 0.266$^{**}$ | -0.098$^{*}$ | -0.032 | 0.111$^{**}$ |
| Attention (AT) | 0.034 | 1 | 0.220$^{**}$ | -0.013 | -0.044 | 0.935$^{**}$ | 0.045 | 0.504$^{**}$ | 0.023 | -0.059 | -0.016 | 0.114$^{**}$ |
| Experience (EX) | - 0.029 | 0.220$^{**}$ | 1 | 0.000 | 0.044 | 0.213$^{**}$ | 0.049 | 0.192$^{**}$ | 0.020 | 0.027 | -0.001 | 0.032 |
| Gender (GE) | - 0.133$^{**}$ | - 0.013 | 0.000 | 1 | -0.033 | -0.021 | -0.087$^{*}$ | -0.030 | -0.318$^{**}$ | 0.016 | -0.017 | -0.013 |
| Group (GR) | 0.030 | - 0.044 | 0.044 | -0.033 | 1 | -0.058 | -0.005 | -0.002 | 0.048 | 0.075 | 0.030 | 0.033 |
| Honesty check (HC) | 0.036 | 0.935$^{**}$ | 0.213$^{**}$ | -0.021 | -0.058 | 1 | 0.050 | 0.310$^{**}$ | 0.048 | -0.083$^{*}$ | -0.007 | 0.132$^{**}$ |
| Income (IN) | - 0.017 | 0.045 | 0.049 | -0.087$^{*}$ | -0.005 | 0.050 | 1 | 0.001 | 0.376$^{**}$ | 0.072 | 0.294$^{**}$ | -0.009 |
| Logic Check (LC) | 0.035 | 0.504$^{**}$ | 0.192$^{**}$ | -0.030 | -0.002 | 0.310$^{**}$ | 0.001 | 1 | -0.009 | 0.031 | -0.022 | 0.139$^{**}$ |
| Marital Status (MS) | 0.266$^{**}$ | 0.023 | 0.020 | -0.318** | 0.048 | 0.048 | 0.376$^{**}$ | -0.009 | 1 | -0.137$^{**}$ | 0.033 | -0.004 |
| Race (RA) | - 0.098$^{*}$ | - 0.059 | 0.027 | 0.016 | 0.075 | -0.083$^{*}$ | 0.072 | 0.031 | -0.137$^{**}$ | 1 | 0.214$^{**}$ | 0.013 |
| Schooling SC | - 0.032 | - 0.016 | - 0.001 | -0.017 | 0.030 | -0.007 | 0.294$^{**}$ | -0.022 | 0.033 | 0.214$^{**}$ | 1 | -0.004 |
| Time Check (TC) | 0.111$^{**}$ | 0.114$^{**}$ | 0.032 | -0.013 | 0.033 | 0.132$^{**}$ | -0.009 | 0.139$^{**}$ | -0.004 | 0.013 | -0.004 | 1 |

$^{*}$ Correlation is significant at the .05 level (2-tailed).

$^{**}$ Correlation is significant at the .01 level (2-tailed).

**Table 5.** Logistic Regression Results.

| Factor | B | S.E. | Waid | df | Sig. | Exp(B) | 95% C.I. for EXP(B) | |
|---|---|---|---|---|---|---|---|---|
| | | | | | | | Lower | Upper |
| Age | 0.129 | 0.195 | 0.438 | 1 | 0.508 | 1.138 | 0.776 | 1.669 |
| Experience | 0.484 | 0.136 | 12.706 | 1 | 0.000 | 1.623 | 1.244 | 2.118 |
| Gender (1) | 0.047 | 0.237 | 0.040 | 1 | 0.842 | 1.048 | 0.659 | 1.667 |
| Income | 0.088 | 0.046 | 3.719 | 1 | 0.054 | 1.092 | 0.999 | 1.195 |
| Marital Status | | | 1.414 | 5 | 0.923 | | | |
| Marital Status (1) | 0.507 | 1.318 | 0.148 | 1 | 0.700 | 1.661 | 0.126 | 21.969 |
| Marital Status (2) | 0.785 | 1.380 | 0.324 | 1 | 0.569 | 2.192 | 0.147 | 32.765 |
| Marital Status (3) | 0.288 | 1.326 | 0.047 | 1 | 0.828 | 1.333 | 0.099 | 17.947 |
| Marital Status (4) | 20.157 | 15112.925 | 0.000 | 1 | 0.999 | 567589914.606 | 0.000 | |
| Marital Status (5) | 20.017 | 22927.392 | 0.000 | 1 | 0.999 | 493655138.500 | 0.000 | |
| Race | | | 1.599 | 3 | 0.660 | | | |
| Race (1) | 0.351 | 0.512 | 0.470 | 1 | 0.493 | 1.421 | 0.521 | 3.876 |
| Race (2) | 0.096 | 0.708 | 0.018 | 1 | 0.892 | 1.101 | 0.275 | 4.411 |
| Race (3) | -0.130 | 0.648 | 0.040 | 1 | 0.841 | 0.878 | 0.246 | 3.128 |
| Schooling | -0.044 | 0.090 | 0.235 | 1 | 0.628 | 0.957 | 0.803 | 1.142 |
| Constant | -3.250 | 1.717 | 3.582 | 1 | 0.058 | 0.039 | | |

## 5 Discussion

Litman et al. [25] describe MTurk as "a constantly evolving marketplace where multiple factors can contribute to data quality". In this work, the attentiveness exhibited by an elite segment of the MTurk worker community was investigated. Specifically, workers holding the coveted 'Master' qualification with an 'Approval Rate' of 98% or more (i.e., completed at least 98% of the tasks they had committed to completing) and had a 'Number of HITS approved' value of 1,000 or more (i.e., the number of these activities exceeded 999). It was conjectured that these characteristics would ensure that this group of workers would be free of behavior reflecting inattentiveness and that this higher level of attentiveness would justify the additional cost attached to using workers holding the 'Master' qualification.

To confirm this hypothesis, an experimental approach was adopted in which participants were asked to complete a simple task involving reading about a hypothetical product and then answering questions on their perceptions of the product. Participant attentiveness was ascertained by using a series of questions originally proposed by Abbey and Meloy [1] and evaluating the amount of time spent on the survey. Surprisingly, the results revealed that over a fifth (22.34%) of the participants were not paying attention, having failed one of the three categories of attention checks. This result could be explained by the work of Chandler et al. [9], who examined the attentiveness of workers with an 'Approval Rate' exceeding 95% and discovered that participants were not always entirely devoted to the current task and were multitasking. In particular, 27% of participants in their sample disclosed that they were with other people while completing the study, 18% were watching television, 14% were listening to music, and 6% were chatting online [9]. This would explain the lack of attention being paid.

We contrast our findings with the work of Peer, Vosgerau, and Acquisti [32], who investigated how attention differs through two experiments. The first experiment compared workers with 'low reputation' (i.e., an 'Approval Rate' below 90%) and 'high reputation' (i.e., an 'Approval Rate' exceeding 95%). The second experiment compared what the authors describe as workers with 'low productivity' (i.e., their 'Number of HITS approved' was less than 100) and 'high productivity' (i.e., their 'Number of HITS approved' was more than 500). At least one attention check question was failed by 33.9% of the 'low reputation' workers, 2.6% of the 'high reputation' workers, 29% of the 'low productivity' workers, and 16.7% of the 'high productivity' workers. Given that we took an even more selective approach than Peer, Vosgerau, and Acquisti [32], our findings are concerning. Our failure rate of 22.34% is closer to what they classify as 'low reputation' (33.9%), and between the 'low productivity' (29%) and 'high productivity' (16.7%) workers. Three possibilities could explain this difference. First, the attention checks we used did not work as expected; second, the high level of compensation was miscommunicated; third, a seasonal influence on attentiveness exists. We explore each of these possibilities in greater detail in the following subsections.

### 5.1 Suitability of the Attention Checks

Abbey and Meloy [1] describe the process of forming attention check questions as a delicate task [1], the concern residing with the nature of these constructs. They give as an example the logic check which comprises questions that “require comprehension of logical relationships”, such as “preferring to eat fruit over paper” [1]. The danger with such questions is that “the more subtle the statement, the less objective these checks can become” [1]. Thus, both the participants’ and the researcher’s responses to a question can be tainted by their interpretations. In the honor check, participants are asked to “reveal their own perceptions of their effort and data validity for the study” [1]. Effectively, the honesty check asks “a respondent to [self] identify the level of effort and attention to detail they [perceive that they] gave to the study and if that effort warrants use of their data” [1]. The weakness of this form of attention check is that respondents may have been paying adequate attention but were overly critical of themselves when submitting their responses. Consequently, they did not respond by selecting the ‘strongly agree’ option. An alternative approach might be to use objective tasks to gauge participant attention, as demonstrated by Aker et al. [3]. However, even these questions have issues. For example, if presented with a mathematical problem, the participant must have the skill to solve the question.

Although our study is not the first to use attention checks with survey research to identify careless (i.e., inattentive) participants [22], the use of attention checks has been questioned by some researchers, as it is believed that this can negatively interact with the survey response quality [22, 44]. Some researchers argue that attention checks should not be used at all [22, 44]. Abbey and Meloy [1] warn that the process to exclude respondents who are not paying attention “can become subjective” in cases where the study is not “largely a replication of known results with expected times, scales, or constructs”. Our attention checks may have been too sensitive. If the criteria for rejection were to fail more than one attention check, the inattentiveness rate would drop to 4.61% (26/564). This rate is closer to what has been found in other studies. More specifically, it is similar to the rate of 4.17% reported by Paolacci, Chandler, and Ipeirotis [31] and closer to the finding of 2.6% by Peer, Vosgerau, and Acquisti [32] (for what they classified as ‘high reputation’ workers with an ‘Approval Rate’ exceeding 95%).

### 5.2 The Effect of Compensation

Another aspect to consider is compensation and its effect on participant attention. In our study, despite offering an extremely high hourly wage to our participants (above $22/hour), we found substantial evidence of inattentiveness, as the high wage did not eliminate the problem of lack of attention exhibited by participants. The magnitude of the compensation we offered can be better understood when it is compared to the median wage for MTurk workers in the United States, which in 2019 was said to be $3.01/hour [16], and the current federal minimum wage of $7.25/hour [43]. Correspondingly, our participants should have been well enticed, and no evidence of inattentiveness should have been discovered. Thus, a high wage does not eliminate the

possibility of having inattentive participants whose work must be discarded. An explanation for this finding might be that participants do not consider the hourly wage but rather the total compensation offered. For example, one may prefer a reward of $0.50/hour if the total compensation of a task were $10 rather than a reward of $20/hour when the total compensation from the task was $1. A tradeoff appears to exist where, as per Aker et al. [3], increasing compensation leads to improved data quality. However, our research suggests that the ability of money to improve attention is limited after a certain point. Additional research is needed to create a better understanding of the marginal effects of wages on participants' attention and identify an optimal point that maximizes attention vis-à-vis compensation.

### 5.3 The Explanation of Seasonality

An alternative explanation of the varying range of inattentiveness exhibited by participants in the studies mentioned above may be found in the work of Chmielewski and Kucker [11], who replicated an experiment four times: the first between December 2015 and January 2016; the second between March 2017 and May 2017; the third between July 2018 and September 2018; and the fourth in April 2019. In their work, the percentage of participants who failed at least one attention check (which they called a "validity indicator") slowly increased from 10.4% to 13.8%, then jumped to 62%, and finally dropped to 38.2%. Given that we collected our data eight months after the conclusion of their final data collection, our inattentiveness rate of 22.34% is not only similar but might indicate a downward trend and possibly a cyclical pattern.

### 5.4 The Irrelevance of User Characteristics

We also attempted to ascertain whether a pattern exists that could help to predict which participants would fail our attention checks. The characteristics of age, gender, income, marital status, race, and schooling were examined, but no relationship was found concerning participant attention. It appears that the lack of attention does not reside with any specific demographic group. Instead, everyone has an equal chance of being inattentive. This outcome is slightly puzzling, as specific demographics have already been linked with participant attention, such as age [26] and culture [27]. The analysis did identify that participants' prior experience of using the technology applied in the study influenced their attention, with those having the most prior experience with the technology exhibiting the greatest attention.

### 5.5 Implications

This work has several noteworthy implications. The first concerns the discovery that participant inattentiveness persists within the population we investigated. This group consisted of MTurk workers with the 'Master' qualification, an 'Approval Rate' of 98% or more, and a 'Number of HITS approved' value of 1,000 or more. Coupled with the high compensation to ensure participants were highly motivated, it is evident that no 'silver bullet' exists that can reliably eliminate the manifestation of participant inattentiveness. Thus, there appears to be no justification in undertaking the additional expense associated with recruiting only participants with the 'Master' qualification. If inattentiveness can be observed under these 'optimal' conditions, this concern cannot be discounted. The fact that there is no one characteristic (i.e., age, education, gender, income, or marital status) that can be used to explain the phenomenon offers minimal hope of an informed intervention. Instead, researchers must vigilantly review participants for inattentiveness and not presume that certain criteria will ensure participants pay attention. Ultimately, the finding highlights the importance of using attention checks to identify inattentive participants and implementing a process to address these occurrences. Specifically, with an inattentiveness rate as high as 22.34%, such a practice would demand "researcher time, funds, and other resources" [11].

A tactic to mitigate the additional cost might be to refuse to compensate any participant who fails to satisfy one or a combination of attention checks. However, this involves challenges. Participants who are refused compensation may object and thus require additional (potentially costly) resources to be invested by the researcher to address those concerns. Participants who have earnestly participated as best they can but failed to produce results that pass the attention check(s) would be unfairly denied compensation. An alternative strategy to withholding payment might be to offer a low rate for participation in studies but offer a bonus for submissions matching a particular pattern. The problem with this approach is that participants may not focus on the research but on producing the illusion that they paid attention. Moreover, this may introduce biases in the responses, as participants may not respond honestly and authentically but rather as they believe the researchers want them to respond.

No simple solution exists. Consequently, to address participant inattentiveness, researchers should consider adjusting their proposals to account for the effort and costs required to identify participants who do not pay attention, address problems arising when addressing their poor performance, and recruit additional participants to replace submissions that must be disregarded.

# 6 References

1. Abbey, J., Meloy, M.: Attention by design: Using attention checks to detect inattentive respondents and improve data quality. Journal of Operations Management. 53–56, 63–70 (2017). https://doi.org/10.1016/j.jom.2017.06.001.
2. Aguinis, H. et al.: MTurk research: Review and recommendations. Journal of Management. 47, 4, 823–837 (2021). https://doi.org/10.1177/0149206320969787.
3. Aker, A. et al.: Assessing crowdsourcing quality through objective tasks. In: Proceedings of the Eighth International Conference on Language Resources and Evaluation (LREC'12). pp. 1456–1461 European Language Resources Association (ELRA), Istanbul, Turkey (2012).

4. Andersen, D., Lau, R.: Pay rates and subject performance in social science experiments using crowdsourced online samples. Journal of Experimental Political Science. 5, 3, 217–229 (2018). https://doi.org/10.1017/XPS.2018.7.
5. Barends, A.J., de Vries, R.E.: Noncompliant responding: Comparing exclusion criteria in MTurk personality research to improve data quality. Personality and Individual Differences. 143, 84–89 (2019). https://doi.org/10.1016/j.paid.2019.02.015.
6. Berry, D.T.R. et al.: MMPI-2 random responding indices: Validation using a self-report methodology. Psychological Assessment. 4, 3, 340–345 (1992). https://doi.org/10.1037/1040-3590.4.3.340.
7. Buhrmester, M. et al.: Amazon's Mechanical Turk: A new source of inexpensive, yet high-quality, data? Perspectives on Psychological Science. 6, 1, 3–5 (2011). https://doi.org/10.1177/1745691610393980.
8. Casler, K. et al.: Separate but equal? A comparison of participants and data gathered via Amazon's MTurk, social media, and face-to-face behavioral testing. Computers in Human Behavior. 29, 6, 2156–2160 (2013). https://doi.org/10.1016/j.chb.2013.05.009.
9. Chandler, J. et al.: Nonnaïveté among Amazon Mechanical Turk workers: Consequences and solutions for behavioral researchers. Behavior Research Methods. 46, 1, 112–130 (2014). https://doi.org/10.3758/s13428-013-0365-7.
10. Chen, J.J. et al.: Opportunities for crowdsourcing research on Amazon Mechanical Turk. Presented at the CHI 2011 Workshop on Crowdsourcing and human computation, https://www.humancomputation.com/crowdcamp/chi2011/papers/chen-jenny.pdf, last accessed 2021/6/9.
11. Chmielewski, M., Kucker, S.C.: An MTurk Crisis? Shifts in data quality and the impact on study results. Social Psychological and Personality Science. 11, 4, 464–473 (2019). https://doi.org/10.1177/1948550619875149.
12. Crump, M.J.C. et al.: Evaluating Amazon's Mechanical Turk as a tool for experimental behavioral research. PloS One. 8, 3, 1–18 (2013). https://doi.org/10.1371/journal.pone.0057410.
13. Difallah, D. et al.: Demographics and dynamics of Mechanical Turk workers. In: Proceedings of the Eleventh ACM International Conference on Web Search and Data Mining. pp. 135–143 Association for Computing Machinery, New York, NY, USA (2018). https://doi.org/10.1145/3159652.3159661.
14. Fort, K. et al.: Amazon Mechanical Turk: Gold mine or coal mine? Computational Linguistics. 37, 2, 413–420 (2011). https://doi.org/10.1162/COLI_a_00057.
15. Goodman, J.K. et al.: Data collection in a flat world: The strengths and weaknesses of Mechanical Turk samples. Journal of Behavioral Decision Making. 26, 3, 213–224 (2013). https://doi.org/10.1002/bdm.1753.
16. Hara, K. et al.: Worker demographics and earnings on Amazon Mechanical Turk: An exploratory analysis. In: Extended Abstracts of the 2019 CHI Conference on Human Factors in Computing Systems. pp. 1–6 ACM Inc., New York, NY, USA (2019). https://doi.org/10.1145/3290607.3312970.

17. Hauser, D.J., Schwarz, N.: Attentive Turkers: MTurk participants perform better on online attention checks than do subject pool participants. Behavior Research Methods. 48, 1, 400–407 (2016). https://doi.org/10.3758/s13428-015-0578-z.
18. Holland, A.: How Estimated Reading Times Increase Engagement with Content, https://marketingland.com/estimated-reading-times-increase-engagement-79830, last accessed 2021/6/9.
19. Horton, J.J., Chilton, L.B.: The labor economics of paid crowdsourcing. In: Proceedings of the 11th ACM Conference on Electronic Commerce. pp. 209–218 ACM Inc., Cambridge, Massachusetts, USA (2010). https://doi.org/10.1145/1807342.1807376.
20. Hydock, C.: Assessing and overcoming participant dishonesty in online data collection. Behavior Research Methods. 50, 4, 1563–1567 (2018). https://doi.org/10.3758/s13428-017-0984-5.
21. Johnson, J.A.: Ascertaining the validity of individual protocols from Web-based personality inventories. Journal of Research in Personality. 39, 1, 103–129 (2005). https://doi.org/10.1016/j.jrp.2004.09.009.
22. Kung, F.Y.H. et al.: Are attention check questions a threat to scale validity? Applied Psychology: An International Review. 67, 2, 264–283 (2018). https://doi.org/10.1111/apps.12108.
23. Laerd Statistics: Binomial Logistic Regression using SPSS Statistics, https://statistics.laerd.com/spss-tutorials/binomial-logistic-regression-using-spss-statistics.php#procedure, last accessed 2020/11/29.
24. Levay, K.E. et al.: The demographic and political composition of Mechanical Turk samples. SAGE Open. 6, 1, (2016). https://doi.org/10.1177/2158244016636433.
25. Litman, L. et al.: The relationship between motivation, monetary compensation, and data quality among US- and India-based workers on Mechanical Turk. Behavior Research Methods. 47, 2, 519–528 (2015). https://doi.org/10.3758/s13428-014-0483-x.
26. Lufi, D., Haimov, I.: Effects of age on attention level: Changes in performance between the ages of 12 and 90. Aging, Neuropsychology, and Cognition. 26, 6, 904–919 (2019). https://doi.org/10.1080/13825585.2018.1546820.
27. Masuda, T.: Culture and attention: Recent empirical findings and new directions in cultural psychology. Social and Personality Psychology Compass. 11, 12, e12363 (2017). https://doi.org/10.1111/spc3.12363.
28. Meade, A.W., Craig, S.B.: Identifying careless responses in survey data. Psychological Methods. 17, 3, 437–455 (2012). https://doi.org/10.1037/a0028085.
29. Okumus, B. et al.: Psychological factors influencing customers' acceptance of smartphone diet apps when ordering food at restaurants. International Journal of Hospitality Management. 72, 67–77 (2018). https://doi.org/10.1016/j.ijhm.2018.01.001.
30. Palan, S., Schitter, C.: Prolific.ac—A subject pool for online experiments. Journal of Behavioral and Experimental Finance. 17, 22–27 (2018). https://doi.org/10.1016/j.jbef.2017.12.004.

31. Paolacci, G. et al.: Running experiments on Amazon Mechanical Turk. Judgment and Decision Making. 5, 5, 411–419 (2010).
32. Peer, E. et al.: Reputation as a sufficient condition for data quality on Amazon Mechanical Turk. Behavior Research Methods. 46, 4, 1023–1031 (2014). https://doi.org/10.3758/s13428-013-0434-y.
33. Pittman, M., Sheehan, K.: Amazon's Mechanical Turk a digital sweatshop? Transparency and accountability in crowdsourced online research. Journal of Media Ethics. 31, 4, 260–262 (2016). https://doi.org/10.1080/23736992.2016.1228811.
34. Rand, D.G.: The promise of Mechanical Turk: How online labor markets can help theorists run behavioral experiments. Journal of Theoretical Biology. 299, 172–179 (2012). https://doi.org/10.1016/j.jtbi.2011.03.004.
35. Rondan-Cataluña, F.J. et al.: A comparison of the different versions of popular technology acceptance models: A non-linear perspective. Kybernetes. 44, 5, 788–805 (2015). https://doi.org/10.1108/K-09-2014-0184.
36. Ross, J. et al.: Who are the crowdworkers? Shifting demographics in Mechanical Turk. In: CHI '10 Extended Abstracts on Human Factors in Computing Systems. pp. 2863–2872 Association for Computing Machinery, New York, NY, USA (2010). https://doi.org/10.1145/1753846.1753873.
37. Rouse, S.V.: A reliability analysis of Mechanical Turk data. Computers in Human Behavior. 43, 304–307 (2015). https://doi.org/10.1016/j.chb.2014.11.004.
38. Salinas-Segura, A., Thiesse, F.: Extending UTAUT2 to explore pervasive Information systems. In: Proceedings of the 23rd European Conference on Information Systems. pp. 1–17 Association for Information Systems, Münster, DE (2015). https://doi.org/10.18151/7217456.
39. Schmidt, G.B., Jettinghoff, W.M.: Using Amazon Mechanical Turk and other compensated crowdsourcing sites. Business Horizons. 59, 4, 391–400 (2016). https://doi.org/10.1016/j.bushor.2016.02.004.
40. Stewart, N. et al.: The average laboratory samples a population of 7,300 Amazon Mechanical Turk workers. Judgment and Decision Making. 10, 5, 479–491 (2015).
41. Stieninger, M. et al.: Factors influencing the organizational adoption of cloud computing: A survey among cloud workers. International Journal of Information Systems and Project Management. 6, 1, 5–23 (2018).
42. Stone, A.A. et al.: MTurk participants have substantially lower evaluative subjective well-being than other survey participants. Computers in Human Behavior. 94, 1–8 (2019). https://doi.org/10.1016/j.chb.2018.12.042.
43. U.S. Department of Labor: Minimum Wage, https://www.dol.gov/general/topic/wages/minimumwage, last accessed 2020/11/25.
44. Vannette, D.: Using Attention Checks in your Surveys May Harm Data Quality, https://www.qualtrics.com/blog/using-attention-checks-in-your-surveys-may-harm-data-quality/, last accessed 2021/01/07.

45. Venkatesh, V. et al.: Consumer acceptance and use of information technology: Extending the Unified Theory of Acceptance and Use of Technology. MIS Quarterly. 36, 1, 157–178 (2012). https://doi.org/10.2307/41410412.
46. Venkatesh, V., Bala, H.: Technology Acceptance Model 3 and a research agenda on interventions. Decision Sciences. 39, 2, 273–315 (2008). https://doi.org/10.1111/j.1540-5915.2008.00192.x.
47. Versta Research: How to Estimate the Length of a Survey, https://verstaresearch.com/newsletters/how-to-estimate-the-length-of-a-survey/, last accessed 2020/04/10.
48. Yang, H.C., Wang, Y.: Social sharing of online videos: Examining American consumers' video sharing attitudes, intent, and behavior. Psychology & Marketing. 32, 9, 907–919 (2015). https://doi.org/10.1002/mar.20826.
49. Yoo, W. et al.: Drone delivery: Factors affecting the public's attitude and intention to adopt. Telematics and Informatics. 35, 6, 1687–1700 (2018). https://doi.org/10.1016/j.tele.2018.04.014.
50. Zack, E.S. et al.: Can nonprobability samples be used for social science research? A cautionary tale. Survey Research Methods. 13, 215–227 (2019).
51. Zimmerman, J. et al.: Field trial of Tiramisu: Crowd-sourcing bus arrival times to spur co-design. In: Proceedings of the SIGCHI Conference on Human Factors in Computing Systems. pp. 1677–1686 Association for Computing Machinery, New York, NY, USA (2011). https://doi.org/10.1145/1978942.1979187.